\documentclass[doublecol]{epl2}

\usepackage{amsmath}
\usepackage{mathtools}
\usepackage{amssymb}
\usepackage{cancel}
\usepackage{amsfonts}
\usepackage[utf8]{inputenc}
\usepackage{graphicx}
\usepackage{xcolor}
\usepackage[colorlinks=true,linkcolor=blue,urlcolor=blue,citecolor=blue]{hyperref}

\begin{document}

\title{Wave function correlations and the AC conductivity of disordered wires beyond the Mott-Berezinskii law}
\shorttitle{AC conductivity of disordered wires}

\author{G. M. Falco\inst{1} \and Andrei A. Fedorenko\inst{2} \and Ilya A. Gruzberg\inst{3}}
\shortauthor{G. M. Falco, A. A. Fedorenko, and I. A. Gruzberg, EPL \textbf{120}, 37004 (2017)}

\institute{
  \inst{1} Amsterdam University of Applied Studies, Weesperzijde 190, 1097 DZ, Amsterdam, the Netherlands\\
  \inst{2} Univ Lyon, ENS de Lyon, Univ Claude Bernard, CNRS, Laboratoire de Physique, F-69342 Lyon, France\\
  \inst{3} Ohio State University, Department of Physics, 191 W. Woodruff Ave, Columbus OH, 43210, USA
}


\date{January 30, 2018}

\abstract{
In one-dimensional disordered wires electronic states are localized at any energy. Correlations of the states at close positive energies and the AC conductivity $\sigma(\omega)$ in the limit of small frequency are described by the Mott-Berezinskii theory. We revisit the instanton approach to the statistics of wave functions and AC transport valid in the tails of the spectrum (large negative energies). Applying our recent results on functional determinants, we calculate exactly the integral over gaussian fluctuations around the exact two-instanton saddle point. We derive correlators of wave functions at different energies beyond the leading order in the energy difference. This allows us to calculate corrections to the Mott-Berezinskii law (the leading small frequency asymptotic behavior of $\sigma(\omega)$) which approximate the exact result in a broad range of $\omega$. We compare our results with the ones obtained for positive energies.
}

\maketitle

\section{Introduction}

One-dimensional (1D) systems have played an important role in developing the theory of coherent quantum transport in disordered solids. Examples include Anderson localization of non-interacting particles in the presence of disorder \cite{Anderson:1958}, the Mott insulating phase in interacting systems \cite{Giamarchi:2003} and the recently discovered many-body localization which takes place in the middle of the spectrum of disordered interacting systems \cite{Vosk:2015}.

In the absence of interactions and decoherence, electrons in a 1D wire are localized at any energy even by a weak random potential.\footnote{In this paper we only consider systems in the unitary class.} Thus, 1D wires lack some features of the higher dimensional systems, such as mobility edges. On the other hand, 1D systems are amenable to powerful non-perturbative methods, such as the phase formalism, which provide access to spectral and localization properties which are much more difficult to obtain in higher dimensions \cite{Lifshitz:1988}. For example, in 1D it is possible to calculate exactly the average density of states (DOS), the localization
length and the Lyapunov exponent, quantities that describe statistics of a single localized wave function.

Less is known about the wave function correlations at different energies and dynamical response functions, such as the finite-frequency (AC) conductivity $\sigma(\omega)$. The mathematical description of such correlations is quite involved and only provided asymptotic analytical expressions exact in the limit of small energy differences \cite{Berezinskii:1973, Gor'kov-I:1983, Gor'kov-II:1983}. An important result of this kind is the behavior of the dissipative AC conductivity $\text{Re} \, \sigma(\omega)$ at low frequencies expressed by the Mott-Berezinskii (MB) formula \cite{Berezinskii:1973, Mott:1968}.

According to the intuitive arguments by Mott, correlations of wave functions at close energies may be described in terms of hybridization of localized states. The leading mechanism for the AC conductivity is the resonant tunneling between pairs of localized states with energies $E-\omega/2$ and $E+\omega/2$ near the Fermi energy $E$. Mott argued that the conductivity is dominated by the tunneling events between states that are separated by the optimal distance (Mott scale) $L_M \sim |\ln \omega|$. This leads to the low-frequency behavior $\text{Re} \, \sigma(\omega) \propto \omega^2 \ln^2 \omega$ in 1D.

Ivanov and co-authors \cite{Ivanov:2012} augmented Mott's arguments by combining them with assumptions about statistics of the localized wave functions and hybridization matrix elements. As a result, the authors quantitatively reproduced the asymptotic features of correlators found in \cite{Gor'kov-I:1983, Gor'kov-II:1983}.

Mott's arguments can be put on a rigorous basis when the resonant states are
deep in the Lifshits tails of the spectrum. Then one can apply the instanton
approach, where the dominant contribution to observables is given by a saddle
point of the action, and subleading contributions come from Gaussain fluctuations
about the saddle point. This method has been used to calculate the average
single-particle Green functions (GFs) and the DOS in systems with Gaussian
white noise \cite{Zittartz:1966, Cardy:1978, Houghton:1979, Brezin:1980, Yaida:2016}
and correlated disorder \cite{John:1984}, systems in magnetic
field \cite{Affleck:1983}, the Lorentz model with repulsive
scatterers~\cite{Luttinger:1983}, systems with bounded non-Gaussian
disorder~\cite{Nieuwenhuizen:1989}, and speckle potentials~\cite{Falco:2015}.

The instanton approach was applied in refs.\cite{Haughton:1980, Hayn:1991}
to the average two-particle dynamic correlation function $S(\omega,x)$ and
the AC conductivity $\sigma(\omega)$ for small $\omega$. These observables
can be written as path integrals which are dominated by non-trivial
two-instanton saddle points. The latter correspond to hybridized states in
the Mott's qualitative picture. Both the saddle points and the fluctuations
around them were found only approximately in ref.~\cite{Haughton:1980},
which led to some inconsistent results. The authors of ref. \cite{Hayn:1991}
found exact two-instanton solutions, but their treatment of fluctuations was
still approximate and restricted only to small frequency $\omega$, reproducing
the asymptotic MB formula. Kirsch et al. \cite{Kirsch:2003} tried to put Mott's argument on a rigorous basis using an expansion in small density of the localizing potential wells, and were able to derive the MB formula as well as asymptotic formulas for wave function correlators in the limit of small energy differences.

Recently we have achieved a significant progress in exact calculations of functional determinants \cite{Falco:2017}. Our results are particularly well suited for applications in instanton calculations. As we have shown, in this case some complicated factors exactly cancel. Using these results, in this paper we calculate exactly the Gaussian fluctuations around the non-trivial two-instanton saddle points and derive the AC conductivity and local DOS correlations applicable in a broad range of frequency $\omega$, and thus going beyond the seminal MB law.

\section{Model}

We consider the model of non-interacting electrons in the presence of disorder in 1D
\begin{align}
H &= - \frac{\hbar^2}{2 m}\nabla^2 + V(x),
& \nabla &\equiv \frac{d}{dx},
\end{align}
where $V(x)$ is a random white noise Gaussian potential:
\begin{align}
\left\langle V(x) \right\rangle &= 0, &
\left\langle V(x) V(x')\right\rangle
&= \gamma \delta(x-x').
\end{align}
We will work in the tails of the spectrum at $E < 0$ and introduce the large dimensionless parameter
\begin{align}
\mathcal{A}_{E} &= \frac{\hbar}{\sqrt{2m}} \frac{|E|^{3/2}}{2\gamma}
\gg 1.
\end{align}
It is convenient to switch to dimensionless quantities by introducing units of length and time:
\begin{align}
\lambda_{E} &= \hbar/\sqrt{2m |E|}, & \tau_{E} &= \hbar/|E|.
\end{align}
These units play a role analogous to the mean free path and time, but unlike the case of $E>0$, they do not depend on the disorder strength, only on $|E|$.

\section{Density of states}

The average DOS at the energy $E$ is obtained from the average GF:
\begin{eqnarray}
&&\!\!\!\!\!\!\!\!\!\!\!\!
\rho(E) = -\frac{\alpha}{\pi}\, \mathrm{Im}\, \mathcal{G}^{\alpha}(0;E),
\ \  \mathcal{G}^{\alpha}(x-x';E)
= \langle G^{\alpha}(x,x';E) \rangle,
\nonumber \\
&& \!\!\!\!\!\!\!\!\!\!\!\! G^{\alpha}(x,x';E) = \langle x|({E+i\alpha 0^+ -H})^{-1}
|x' \rangle ,
\label{eq:Green-fun-1}
\end{eqnarray}
where $\alpha=\pm1$ distinguishes retarded and advanced GFs. We perform the disorder average using the super\-sym\-metry method \cite{Efetov:book}. We introduce a supervector $\Phi(x) = (\phi(x), \chi(x))^T$ and its conjugate $\Phi^\dagger(x) = (\phi^*(x), \bar{\chi}(x))$, where $\phi = \phi_x + i\phi_y$ is a complex bosonic field, and  $\chi$, $\bar{\chi}$ are two fermionic fields. Upon rotation of the fields $\Phi \to \sqrt{-i \alpha} \Phi$ in the complex plane \cite{Hayn:1991}, the average GF and the action in terms of dimensionless variables become
\begin{align}
&\mathcal{G}^{\alpha}(x-x';E) =
-\frac{|E|}{2\gamma}
\int \!\! \mathcal{D}\Phi \, \chi(x) \bar{\chi}(x') e^{-\mathcal{S}}, \label{eq:path-integral}
\\
&\mathcal{S} = {\cal A}_E \int \!\! dx \, \Big[\Phi^\dagger \big(1 - \nabla^2 \big)\Phi - \frac{1}{4} (\Phi^\dagger \Phi)^2 \Big].
\end{align}
The corresponding classical equations of motion are
\begin{align}
\Big(1 - \nabla^2 - \frac{1}{2} |\phi|^2 \Big) \phi &= 0,
& \chi &= 0.
\end{align}
The trivial solution $\phi=0$ does not contribute to the imaginary part of the GF and thus to the DOS. A nontrivial, one-instanton solution that we need is
\begin{align}
\label{eq:saddle-point-solution-density}
\phi_{\mathrm{cl}}(x-x_0,\theta) &=  e^{i\theta} \varphi(x-x_0), & \varphi(x) &= 2/\cosh x.
\end{align}
The parameters $x_0$, $\theta$ describe translations in real space and rotations in the plane $(\phi_x, \phi_y)$. The action of the one-instanton solution $\phi_{\mathrm{cl}}$ does not depend on $x_0$ and $\theta$:
\begin{align}
\mathcal{S}_1(E) = \frac{16}3\mathcal{A}_E
= \frac{8}{3} \frac{\hbar}{\sqrt{2m}} \frac{|E|^{3/2}}{\gamma}.
\end{align}

We now substitute $\phi = \phi_{\mathrm{cl}} + \rho$ into the action and expand to second order in fluctuations $\rho$:
\begin{align} \label{eq:S0+dS}
\mathcal{S} &= \mathcal{S}_1 + \mathcal{A}_E \int \!\! dx \big(\rho_x O_d \rho_x + \rho_y O_s \rho_y + \bar{\chi} O_s \chi \big),
\end{align}
where the fluctuation operators are
\begin{align}
O_\nu =  - \nabla^2 + 1 - 2 C_\nu/\cosh^2(x-x_0),
\end{align}
where $\nu=s, d$ (shallow and deep), $C_s=1$, and $C_d=3$. Gaussian integrals over $\rho$ and $\chi$ give determinants of the operators $O_\nu$. However, these operators have zero modes due to the fact that any particular choice of $x_0$ and $\theta$ breaks symmetries of the action. Whenever a broken symmetry is described by a parameter $\zeta_i$,  the function $\psi_i(x) = \partial_{\zeta_i} \phi_{\mathrm{cl}}$ is a zero mode of one of the fluctuation operators. For instance, the lowest eigenvalue of $O_s$, ${\lambda}_1^s = 0$, is related to breaking the rotation invariance in the $(\phi_x, \phi_y)$ plane. The spectrum of $O_d$ starts with a negative eigenvalue $\lambda_1^d < 0$, followed by the zero eigenvalue $\lambda_2^d=0$ related to breaking the translation invariance. The zero modes of $O_s$ and $O_d$ are
\begin{align}
\psi_s(x-x_0,\theta) &= \partial_\theta \phi_{\mathrm{cl}} = ie^{i\theta} \varphi(x-x_0),
\\
\psi_d(x-x_0,\theta) &= \partial_{x_0} \phi_{\mathrm{cl}}
= -e^{i\theta} \varphi'(x-x_0).
\end{align}

We now separate the negative and zero modes and perform the Gaussian integration of eq.~(\ref{eq:path-integral}) with the action~(\ref{eq:S0+dS}). This yields, formally,
\begin{align}
\mathcal{G}^{\alpha}(x-x';E) &= - \frac{|E|}{2\gamma} e^{-\mathcal{S}_1}
{\Lambda}_1^s {\Lambda}_1^d {\Lambda}_2^d
\Big(\frac{\text{Det}'\mathcal{A}_{E} O_s}{\text{Det}''\mathcal{A}_{E} O_d} \Big)^{1/2}
\nonumber \\
& \quad \times
\varphi(x-x_0) \varphi(x'-x_0)/\langle \varphi | \varphi \rangle,
\label{eq:DOS-res-1}
\end{align}
where $\langle f_1 | f_2 \rangle = \int dx f_1^*(x) f_2(x) $, and $\text{Det}'$ ($\text{Det}''$) stands for a functional determinant with excluded zero (negative and zero) eigenvalues. The contributions of the excluded modes are denoted by ${\Lambda}_1^s$, ${\Lambda}_1^d$ and ${\Lambda}_2^d$. The last factor in eq.~(\ref{eq:DOS-res-1}) comes from the fermionic integral in eq. (\ref{eq:path-integral}) with the action (\ref{eq:S0+dS}).

The contributions ${\Lambda}_1^s$ and ${\Lambda}_2^d$ from the zero modes are computed by introducing the so-called collective coordinates \cite{Langer:1967, Christ:1975, Coleman:1985}. When there are $n$ zero modes $\psi_i$, the collective coordinates are the parameters $\zeta_i$ describing broken symmetries, and the relevant contribution
\begin{align}
\Lambda_{1...n} &= \frac1{\pi^{n/2}} \int \!\! \prod_{i=1}^{n} d\zeta_i  \sqrt{\det \langle \psi_i | \psi_j  \rangle}
\label{eq:contr-zero-1}
\end{align}
involves the determinant of the $n\times n$ matrix whose elements are overlaps of the zero modes $\langle \psi_i | \psi_j  \rangle$. In our case this formula gives
\begin{align}
\Lambda_1^s &= 2 \sqrt{\pi \langle \varphi | \varphi \rangle}, &
\Lambda_2^d  &= \frac{1}{\sqrt{\pi}} \sqrt{\langle \varphi' | \varphi' \rangle} \int \!\! d x_0.
\end{align}

The contribution ${\Lambda}_0^d$ from the negative mode can be computed by an analytic continuation \cite{Langer:1967, Hayn:1991}. We need to rotate the integration contour toward the saddle point in the direction  $\propto \sqrt{-i\alpha}$. At the saddle point the contour turns by $-\alpha \pi/2$ and goes down the valley from the saddle point. One ends up integrating only one half of a Gaussian peak, giving a factor $1/2$ and a phase that depends on $\alpha$:
\begin{align}
\Lambda_1^d &= \int_{-\infty}^{\infty} \!\! \frac{d\zeta}{\sqrt{\pi}} e^{-\lambda_0^d \zeta^2}
\to i\alpha \!\! \int_{0}^{\infty} \!\! \frac{d\zeta}{\sqrt{\pi}} e^{\lambda_0^d\zeta^2} =
\frac{i \alpha}{2 \sqrt{- \lambda_0^d}}.
\label{eq:Lambda0d}
\end{align}
Combining this contribution~(\ref{eq:Lambda0d}) with the ratio
of determinants in eq.~(\ref{eq:DOS-res-1}) we obtain
\begin{align}
\Lambda_1^d \Big(\frac{\text{Det}' \mathcal{A}_{E} O_s}{\text{Det}''\mathcal{A}_{E} O_d} \Big)^{1/2}
= \frac{1}{2} \Big(\frac{\text{Det}' O_s}{\text{Det}' O_d}\Big)^{1/2}.
\label{eq:ratio-2-det-1}
\end{align}
Here $\text{Det}' O_d < 0$, and we need to choose the phase according to eq. (\ref{eq:Lambda0d}).

Calculation of the functional determinants can be done explicitly (see e.g. \cite{Falco:2015}) but in the subsequent study of correlations of wave functions similar explicit calculations will be impossible. Thus, we use results of ref. \cite{Falco:2017} (generalizing those of \cite{Gelfand:1960, Tarlie:1995}), where we have shown that when an $n\times n$ matrix Schr\"odinger operator $O$ defined on the interval $x \in (a,b)$ with homogeneous boundary conditions has $n$ zero modes $\psi_i(x)$ (vectors with components $\psi_{ij}(x)$, $i,j = 1,..,n$), its functional determinant with excluded zero eigenvalues is equal to
\begin{align}
\text{Det}' {O} = (-1)^n \frac{\det \langle \psi_i|\psi_j \rangle}{\det \psi_{ij}'(a) [\det \psi_{ij}'(b)]^*}.
\label{eq:det-Omega-NN}
\end{align}
This equation is formal and needs to be used in a ratio of two determinants. For the ratio in eq. (\ref{eq:ratio-2-det-1}) we obtain
\begin{align}
\frac{\text{Det}' O_s}{\text{Det}' O_d} =
\frac{\langle \varphi | \varphi \rangle}{\langle \varphi' | \varphi' \rangle}
\lim\limits_{R\to \infty} \frac{ \varphi''(-R) \varphi''(R) }{ \varphi'(-R) \varphi'(R) }
= - \frac{\langle \varphi | \varphi \rangle}{\langle \varphi' | \varphi' \rangle},
\end{align}
where in the limit it is sufficient to use the asymptotic form of the solution $\varphi(x)$.

When all the factors are combined, the overlaps of the zero modes cancel, as they should \cite{Falco:2017}, and we get
\begin{align}
\mathcal{G}^{\alpha}(x-x';E) &= -i\alpha \frac{4|E|}{\gamma}
e^{-\mathcal{S}_1} \frac{x-x'}{\sinh(x-x')}.
\end{align}
For the DOS we get a well-known expression
\begin{align}
\rho(E) = \frac{4|E|}{\pi \gamma}
\exp\bigg(\!\!-\frac{8}{3} \frac{\hbar}{\sqrt{2m}} \frac{|E|^{3/2}}{\gamma} \bigg).
\label{eq:DOS-result}
\end{align}

\section{Correlation functions}

Statistics of localized states at two energies $E_{1,2} = E \mp \hbar\omega/2$, are characterized by the two-point local DOS correlation function \cite{Gor'kov-I:1983}
\begin{align}
R(\omega,x) &= \rho_m^{-2} \Big\langle \sum_{kl}
\delta(E_k-E_1) \delta(E_l-E_2)
\nonumber \\
& \quad \times
|\psi_k(0)|^2 |\psi_l(x)|^2 \Big\rangle,
\label{eq:DOS-cor-fun-1}
\end{align}
and (the real part of) the dynamic correlation function~\cite{Gor'kov-II:1983}
\begin{align}
S(\omega,x) &= \rho_m^{-2} \text{Re} \Big\langle \sum_{kl}
\delta(E_k-E_1) \delta(E_l-E_2)
\nonumber \\
& \quad \times
\psi_k(0) \psi_k^*(x) \psi_l(x) \psi_l^*(0) \Big\rangle.
\label{eq:DOS-cor-fun-2}
\end{align}
Here  $\rho_m^{2} = \rho(E_1) \rho(E_2)$. Functions~(\ref{eq:DOS-cor-fun-1}) and (\ref{eq:DOS-cor-fun-2}) can be calculated from disorder averages of products of GFs:
\begin{align}
& {\cal G}^{\alpha_1 \alpha_2}(x_1,x_1';x_2,x_2')
= \Big\langle \prod_{a=1}^{2} G^{\alpha_a}(x_a,x_a';E_a)
\Big\rangle
\label{eq:cor-fun-RA}
\end{align}
by means of
\begin{align}
R(\omega, x) &= -\frac{1}{4\pi^2 \rho_m^2}
\sum_{\alpha_1, \alpha_2} \alpha_1 \alpha_2
{\cal G}^{\alpha_1 \alpha_2}(0,0,x,x),
\label{eq:DOS-cor-fun-10}
\\
S(\omega,x) &= -\frac{1}{4\pi^2 \rho_m^2}
\text{Re} \! \sum_{\alpha_1, \alpha_2} \alpha_1 \alpha_2
{\cal G}^{\alpha_1 \alpha_2}(0,x,x,0).
\label{eq:DOS-cor-fun-20}
\end{align}

Assuming that both energies are in the tail of the spectrum, $E_a < 0$, let us introduce the following notation:
\begin{align}
E_{a} &= -|E|k_{a}^2, &
k_{1,2} &= \sqrt{1 \pm \bar{\omega}}, &
{\bar \omega} &= \hbar \omega/2|E|.
\label{E-1-2}
\end{align}
The supersymmetry representation for ${\cal G}^{\alpha_1 \alpha_2}$ requires two superfields $\Phi_a = (\phi_a, \chi_a)$, $a = 1,2$. After rotations of the fields in the complex plane we have
\begin{align*}
&{\cal G}^{\alpha_1 \alpha_2}(x_1,x_1';x_2,x_2')
= \frac{|E|^2}{4 \gamma^2}
\int \!\! \prod_{a} {\cal D} \Phi_a \chi_a(x_a) \bar{\chi}_a(x'_a) e^{-\mathcal{S}},
\nonumber \\
&\mathcal{S} = {\cal A}_E \int dx
\Big[\sum_{a} \Phi_a^\dagger \big(k_a^2 - \nabla^2 \big)\Phi_a
- \frac{1}{4} \Big(\sum_{a} \Phi_a^\dagger \Phi_a \Big)^2 \Big].
\end{align*}
The bosonic saddle point equation is a two-component non-linear Schr\"{o}dinger equation
\begin{align}
\label{eq:classical}
\Big[k_{a}^2 - \nabla^2 - \frac{1}{2}(|\phi_1|^2 + |\phi_2|^2)\Big]\phi_{a} &= 0, & a &= 1,2.
\end{align}
Equation (\ref{eq:classical}) has the trivial solution $(\phi_{1,2}=0)$, two one-instanton solutions $(\phi_1 \neq 0$, $\phi_2 = 0)$, and $(\phi_1 = 0$, $\phi_2 \neq 0)$, and a two-instanton solution with $(\phi_{1,2} \neq 0)$. The trivial
and the one-instanton solutions do not contribute to the functions $R$ and $S$, which are determined by the two-instanton solutions.

The exact two-instanton solutions found in ref. \cite{Hayn:1991}
can be written as $\phi_{a,\text{cl}}(x-x_0) = e^{i\theta_a} \varphi_a(x-x_0)$, where
\begin{align}
&\begin{pmatrix} \varphi_1(x) \\ \varphi_2(x) \end{pmatrix}
= \begin{pmatrix}
    \cos \theta(x) & \sin \theta(x) \\
    \sin \theta(x) & -\cos \theta(x)
  \end{pmatrix}
\begin{pmatrix} \varphi_L(x) \\ \varphi_R(x) \end{pmatrix},
\label{two-instanton-mixture} \\
&\varphi_L(x) = \frac{4 k_1 e^{-x_1}}{1 + e^{-2x_1} + e^{2x_2}},
\quad
\varphi_R(x) = \frac{4 k_2 e^{x_2}}{1 + e^{-2x_1} + e^{2x_2}},
\nonumber \\
& 2 x_{1,2}(x) = (k_1 + k_2)x \pm [D_0 - \ln \sin 2\theta(x)],
\nonumber \\
& D_0 = \ln \frac{2(k_1 + k_2)}{k_1 - k_2},
\quad
\cot \theta(x) = e^{(k_1 - k_2)x + f_0}.
\end{align}
These solutions contain four free parameters $\theta_1, \theta_2, x_0, f_0$, so we expect to have four zero modes in the fluctuation spectrum. The parameter $f_0$ determines the distance
\begin{align}
D = D_0 + \ln \cosh f_0
\label{D-theta0}
\end{align}
between the left and right ``instantons'' $\varphi_L$ and $\varphi_R$, whose minimal value $D_0$ ($\sim \ln(4/\bar{\omega})$ for $\bar{\omega} \ll 1$) plays the role of the Mott scale $L_M$. The action of a two-instanton solution does not depend on the free parameters:
\begin{align}
\mathcal{S}_2 = \mathcal{S}_1(E_1) + \mathcal{S}_1(E_2)
= \frac{16}{3} (\mathcal{A}_{E_1} + \mathcal{A}_{E_2}).
\label{S_2}
\end{align}

We now introduce the fluctuation fields $\rho_x = (\rho_{1x}, \rho_{2x})$, $\rho_y = (\rho_{1y}, \rho_{2y})$, $\chi^\dagger = (\bar{\chi}_1, \bar{\chi}_2)$,
and expand the action around the saddle point:
\begin{align}
\mathcal{S} = \mathcal{S}_2 + \mathcal{A}_E \int \! dx \, [\rho_x O_x \rho_x + \rho_y O_y \rho_y + \chi^\dagger O_y \chi],
\end{align}
where the fluctuation $2\times 2$ matrix operators are
\begin{align}
O_y &= -\nabla^2 -\frac{1}{2} (\varphi_1^2+\varphi_2^2) +
\begin{pmatrix} k_1^2  & 0 \\ 0 & k_2^2 \end{pmatrix},
\nonumber \\
O_x &= O_y - \begin{pmatrix}
\varphi_1^2 & \varphi_1 \varphi_2 \\
\varphi_1 \varphi_2 & \varphi_2^2
\end{pmatrix}.
\end{align}

The operator $O_x$ has two negative modes $\psi^x_{1}$ and $\psi^x_{2}$ (whose explicit form we do not need) and two zero modes
\begin{align}
\psi^x_{3}(x) &=
\begin{pmatrix} \varphi_1'(x) \\ \varphi_2'(x) \end{pmatrix}, &
\psi^x_{4}(x) &= \begin{pmatrix} \partial_{f_0}\varphi_1(x) \\ \partial_{f_0}\varphi_2(x) \end{pmatrix},
\end{align}
related to the translation invariance and to changing the distance $D$. The operator $O_y$ has two zero modes
\begin{align}
\psi^y_{1}(x) &= \begin{pmatrix} \varphi_1(x) \\ 0 \end{pmatrix}, &
\psi^y_{2}(x) &= \begin{pmatrix} 0 \\ \varphi_2(x) \end{pmatrix},
\end{align}
related to the rotation invariance with respect to $\theta_a$.

A formal Gaussian integration gives
\begin{align}
\mathcal{G}^{\alpha_1 \alpha_2} &=
\frac{|E|^2}{4 \gamma^2} e^{-\mathcal{S}_2}
\Lambda^{x}_{12} \Lambda^{x}_{34} \Lambda^{y}_{12}
\Big( \frac{\text{Det}' \mathcal{A}_E O_y}{\text{Det}'' \mathcal{A}_E O_x} \Big)^{1/2}
\nonumber \\
& \quad \times
\prod_a \frac{\varphi_a (x_a - x_0) \varphi_a(x_a' - x_0)}{\langle\varphi_a|\varphi_a \rangle}.
\label{eq:res11}
\end{align}
Equation~(\ref{eq:contr-zero-1}) gives the contributions of zero modes:
\begin{align}
\Lambda^{x}_{34} &=
\frac1{\pi} \int \!\! df_0 \int \!\! dx_0 \,
\sqrt{\langle \psi^{x}_3| \psi^{x}_3 \rangle
\langle \psi^{x}_4| \psi^{x}_4 \rangle
- |\langle \psi^{x}_3| \psi^{x}_4 \rangle|^2},
\nonumber \\
\Lambda^{y}_{12} &=
4\pi \sqrt{\left\langle \varphi_1 | \varphi_1 \right\rangle
\left\langle \varphi_2 | \varphi_2 \right\rangle}.
\end{align}
Similarly to eq. (\ref{eq:ratio-2-det-1}) we now have
\begin{align}
\Lambda^{x}_{12} \Big(\frac{\text{Det}' \mathcal{A}_{E} O_y}{\text{Det}''\mathcal{A}_{E} O_x} \Big)^{1/2}
= -\frac{\alpha_1 \alpha_2}{4} \Big(\frac{\text{Det}' O_y}{\text{Det}' O_x}\Big)^{1/2}.
\end{align}
To compute this determinant ratio we use
eq.~(\ref{eq:det-Omega-NN}):
\begin{align}
\frac{\text{Det}' O_y}{\text{Det}' O_x} =
\frac{\langle \varphi_1 | \varphi_1 \rangle
\langle \varphi_2 | \varphi_2 \rangle}{\langle \psi^{x}_3| \psi^{x}_3 \rangle
\langle \psi^{x}_4| \psi^{x}_4 \rangle
- |\langle \psi^{x}_3| \psi^{x}_4 \rangle|^2} F^2(\bar{\omega}),
\end{align}
where $F(\bar{\omega})$  is easily computed from the asymptotic behavior of the solutions (\ref{two-instanton-mixture}) at infinity:
\begin{align}
F(\bar{\omega}) &= \lim\limits_{R \to \infty}\frac{
{\varphi_2}''(R) {\partial_{f_0}{\varphi_1}'}(R)-{\varphi_1}''(R)
{\partial_{f_0}{\varphi_2}'}(R)}{{\varphi}'_1 (R) {\varphi}'_2 (R)}
\nonumber \\
&= \frac{k_1 + k_2}{2}
= \frac{\sqrt{1+\bar{\omega}} + \sqrt{1-\bar{\omega}}}{2}.
\end{align}

Using eqs.~(\ref{eq:DOS-result}), (\ref{E-1-2}), and (\ref{S_2}), we have
\begin{align}
\frac{|E|^2}{4\gamma^2} e^{-\mathcal{S}_2}
= \frac{\pi^2 \rho_m^2}{2^6 k_1^2 k_2^2}
= \frac{\pi^2 \rho_m^2}{2^6 (1 - \bar{\omega}^2)}.
\end{align}
Collecting all factors we arrive at
\begin{eqnarray}
&&\!\!\!\!\!\!\!\!\!\! {\cal G}^{\alpha_1 \alpha_2}
= -\alpha_1 \alpha_2 \pi^2 \rho_m^2 Y(\bar{\omega})
Q,
\ \  Y(\bar{\omega}) = \frac{\sqrt{1+\bar{\omega}}
+ \sqrt{1-\bar{\omega}}}{2(1 - \bar{\omega}^2)},
\nonumber \\
&& \!\!\!\!\!\!\!\!\!\!  Q(x_1,x_1';x_2,x_2') = \frac{1}{2^6} \iint \!\! d f_0 d x_0 \,
\varphi_1(x_1 - x_0) \varphi_1(x_1' - x_0)
\nonumber \\
&& \qquad \qquad \times
\varphi_2(x_2 - x_0) \varphi_2(x_2' - x_0).
\label{eq:funY-4}
\end{eqnarray}
Substituting this into eqs.~(\ref{eq:DOS-cor-fun-10}) and (\ref{eq:DOS-cor-fun-20}), we finally obtain
\begin{align}
R(\omega,x) &= Y(\bar{\omega}) Q_R(x),
\quad
S(\omega,x) = Y(\bar{\omega}) Q_S(x),
\label{R-Q-S-Q}
\\
Q_R(x) &= Q(0,0,x,x), \qquad  Q_S(x) = Q(0,x,x,0).
\label{eq:DOS-cor-fun-40}
\end{align}

\begin{figure}
\centering
\includegraphics[width=0.9\columnwidth]{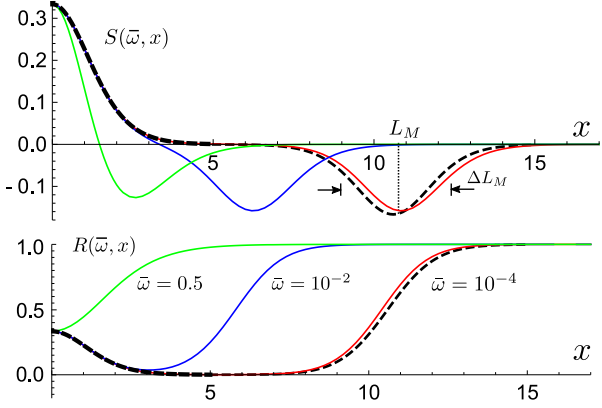}
\caption{The dynamic $S$ and the local DOS $R$ correlation functions.
Solid curves show the numerically exact correlators evaluated for three values of $\bar{\omega}$. The dashed curves near the origin are plots of the exact equation~(\ref{R-S-zero-omega}), and the ones near the Mott scale $L_M$ are plots of the approximations (\ref{Q_R-II}) and (\ref{Q_S-II}).}
\label{fig:corr-fun}
\end{figure}

Equations~(\ref{eq:funY-4})-(\ref{eq:DOS-cor-fun-40}) allow us to establish
some exact properties of the functions $R$ and $S$ such
as $R(\omega,x=0) = S(\omega,x=0) = 1/3$, independently of $\omega$. This property can be established only if the functional determinants are calculated exactly for all $\bar{\omega}$. We also find $R(\omega, x \to \infty) = 1$ as expected from the clustering property of the correlator at large distances. Finally, in the limit $\bar{\omega} \to 0$ we find
\begin{align}
R(0, x) = S(0, x)  = Q_0(x)
\equiv \frac{x \coth x - 1}{\sinh^2 x}.
\label{R-S-zero-omega}
\end{align}
This function describes the peak near $x=0$ shown in fig.~\ref{fig:corr-fun} by a black dashed curve.

\begin{figure}
\centering
\includegraphics[width=0.9\columnwidth]{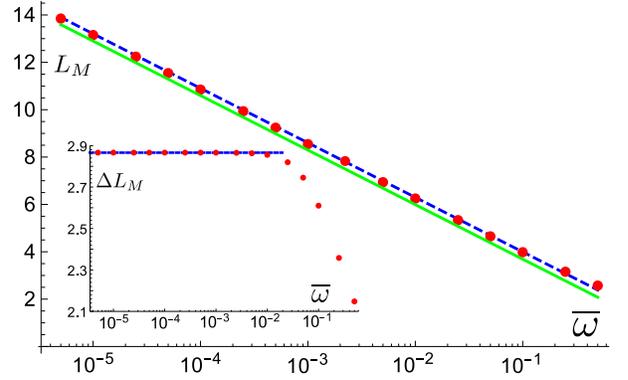}
\caption{The Mott scale $L_M$ as a function of $\bar{\omega}$. The red dots are numerically exctracted minima of $S(\bar{\omega},x)$. The solid line is $\ln (4/\bar{\omega})$, and the dashed line is $D_0 - \ln 2 + 1 \approx \ln(2e/\bar{\omega})$. Inset: the width of the negative peak of $S(\bar{\omega},x)$ as a function of $\bar{\omega}$.}
\label{fig:mott-scale}
\end{figure}

Numerical evaluation of eq.~(\ref{eq:funY-4}) gives the plots of the functions $R$ and $S$ shown in fig.~\ref{fig:corr-fun} for three values of $\bar{\omega}$. The features at the Mott scale (a step in $R$ and a negative bump in $S$) have widths that are independent of $\bar{\omega}$ for $\bar{\omega} \ll 1$, see fig.~\ref{fig:mott-scale} which shows the position $L_M$ of the negative peak of $S(\omega, x)$ and
its width at half maximum $\Delta L_M$ as functions of $\bar{\omega}$.

Analytical expressions for $Q_C$ ($C = R,S$) can be obtained for $\bar{\omega} \ll 1$. In this case one
can neglect the prefactor $Y(\bar{\omega})$, and represent the corrections to the central peak and the features at the Mott scale similar to ref. \cite{Ivanov:2012}:
\begin{align}
Q_C(x) = Q_0(x) + \delta Q_C(x) + Q_{CM}(x).
\end{align}
When $\bar{\omega} \ll 1$, in the leading order the mixing angle $\theta(x)$ in eq. (\ref{two-instanton-mixture}) becomes constant ($\cot \theta_0 = e^{f_0}$), and $\varphi_{L,R}(x)$ become instantons (\ref{eq:saddle-point-solution-density}) separated by $D \gg 1 $ \cite{Hayn:1991}:
\begin{align}
\tilde{\varphi}_L(x) &= \varphi(x+D/2), &
\tilde{\varphi}_R(x) &= \varphi(x-D/2). \label{eq:new}
\end{align}
We substitute these approximate solutions to eq. (\ref{eq:funY-4}), and denote the result by $\tilde{Q}$.
Then the integral over $x_0$ becomes elementary, and after changing the
integration variable $f_0$ to $D$ we obtain for $x>0$
\begin{align}
\delta \tilde{Q}_R(x) &= - 4 \int_{D_0}^{\infty} \!\! dD \, w_1(D-D_0)I(x,D),
\label{delta-tilde-QR}
\\
\tilde{Q}_{RM}(x) &= \int_{D_0}^{\infty} \!\! dD \, w_2(D-D_0)
Q_0(x-D),
\label{tilde-QR}
\\
\delta \tilde{Q}_S(x) &= \int_{D_0}^{\infty} \!\! dD \, w_3(D-D_0) I(x,D),
\label{delta-tilde-QS}
\\
\tilde{Q}_{SM}(x) &= -\int_{D_0}^{\infty} \!\! dD \, w_1(D-D_0) Q_0(x-D),
\label{tilde-QS}
\end{align}
where the tilde indicates that the approximation~(\ref{eq:new}) has been used,
and we have defined the functions
\begin{align}
I(x,D) &= 2 \frac{D \coth D - x \coth x}{\cosh 2D - \cosh 2x},
\quad
w_1(D) = \frac{e^{-D}}{\sqrt{e^{2D} - 1}},
\nonumber \\
w_2(D) &= \frac{2 - e^{-2D}}{2\sqrt{1 - e^{-2D}}},
\quad
w_3(D) = 2 \sqrt{1 - e^{-2D}}.
\end{align}
For $D \gg 1$ and $x  \ll D$ the function $I(x,D)$ can be well approximated by
\begin{align}
I(x,D) &\approx 4 e^{-2D} (D -  x \coth x),
\end{align}
and with this approximation we have
\begin{align}
\delta \tilde{Q}_R(x) &\approx -\frac{16}{3} e^{-2D_0}
(D_0 - x \coth x + 5/6 - \ln 2),
\\
\delta \tilde{Q}_S(x) &\approx \frac{8}{3} e^{-2D_0}
(D_0 - x \coth x  + 4/3 - \ln 2).
\end{align}

The function $w_2(D)$ approaches its asymptotic value of $1$ very rapidly.
Replacing $w_2$ by 1 gives the approximation
\begin{align}
\tilde{Q}_{RM}(x) &\approx I_1(D_0+x) + I_1(D_0-x),
\label{Q_R-II}\\
I_1(x) &= \frac{1 - \coth x}{2} + \frac{x}{2 \sinh^2 x}.
\end{align}
The function $w_1(D)$ is peaked at $D=0$. A reasonable approximation is to replace it by a delta function $\delta(D)/2$:
\begin{align}
\tilde{Q}_{SM}(x) &\approx - \frac{1}{2} \big[Q_0(x-D_0) + Q_0(x+D_0)\big].
\label{Q_S-II}
\end{align}
The approximations (\ref{Q_R-II}) and (\ref{Q_S-II}) are shown in fig. \ref{fig:corr-fun} by dashed lines for $\bar{\omega} = 10^{-4}$.

\section{AC conductivity}

\begin{figure}
\centering
\includegraphics[width=0.9\columnwidth]{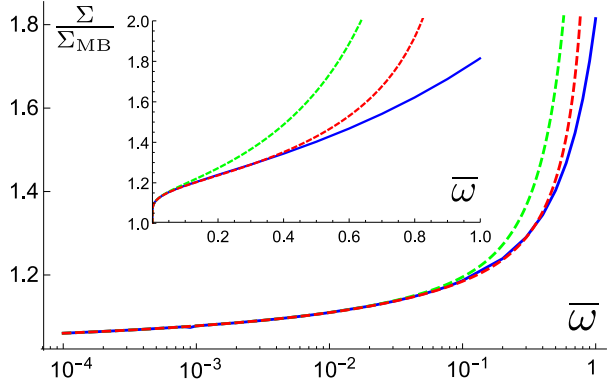}
\caption{The ratio $\Sigma(\bar{\omega})/\Sigma_{\mathrm{MB}}(\bar{\omega})$ as a function of $\ln \bar{\omega}$ (main panel) and $\bar{\omega}$ (inset). The exact formula~(\ref{eq:res-Z-1}) is used for the blue solid curve, the approximation (\ref{Sigma1}) for the green dashed curve, and the sum of (\ref{Sigma1}) and (\ref{Sigma2}) for the red dashed curve.}
\label{fig:Ratios}
\end{figure}

The real part of the AC conductivity $\sigma(\omega)$ is obtained from $S(\omega, x)$ as
\begin{align}
{\rm Re}\, \sigma(\omega) &= \frac{\pi \hbar^4 e^2 \rho_m^2}{2(2m |E|)^{3/2}} \omega^2 \Sigma(\bar{\omega}),
\label{sigma}
\\
\Sigma(\bar{\omega}) &= - Y(\bar{\omega}) \int \!\! dx \, x^2 Q_S(x).
\label{eq:res-Z-1}
\end{align}
This expression is valid for any $E$ in the tail, but arbitrary $0 \leqslant \bar{\omega} < 1$. It can be evaluated with a various degree of accuracy. For $\bar{\omega} \ll 1$ we can replace $Q_S$ by $\tilde{Q}_S$. If we neglect $Y(\bar{\omega})$ in front of the integral in eq. (\ref{eq:res-Z-1}), then neglect $\delta \tilde{Q}_S$, and use eq. (\ref{Q_S-II}) for $\tilde{Q}_{SM}$, we get an approximation for $\Sigma(\bar{\omega})$:
\begin{align}
\Sigma(\bar{\omega}) \approx D_0^2 \approx \ln^2 \frac{4}{\bar{\omega}} \equiv \Sigma_{\text{MB}}(\bar{\omega}).
\end{align}
The last expression is what is called the Mott formula in ref. \cite{Hayn:1991}. Deviations of the numerically exact $\Sigma(\bar{\omega})$ from $\Sigma_{\text{MB}}(\bar{\omega})$ are demonstrated in fig.~\ref{fig:Ratios}, where the solid curve is the plot of the ratio $\Sigma/\Sigma_{\text{MB}}$.

A much better approximation is obtained if we keep $Y(\bar{\omega})$, substitute eqs.~(\ref{delta-tilde-QS}) and (\ref{tilde-QS}) into (\ref{eq:res-Z-1}), and perform the $x$ integration. This gives
$\Sigma(\bar{\omega}) \approx \Sigma_1(\bar{\omega}) + \Sigma_2(\bar{\omega})$, where
\begin{align}
\Sigma_1(\bar{\omega}) &= 2 Y(\bar{\omega}) \int_{D_0}^{\infty} \!\! dD \, w_1(D-D_0) D^2
\nonumber \\
& = Y(\bar{\omega})\big[(D_0 - \ln 2 + 1)^2 + 1 - \pi^2/12\big],
\label{Sigma1}
\\
\Sigma_2(\bar{\omega}) &=
- \frac{Y(\bar{\omega})}{6} \int_{D_0}^{\infty} \!\! dD \, w_3(D-D_0) \frac{D^2(D^2 + \pi^2)}{\sinh^2 D}.
\end{align}
The last integral can be evaluated after we approximate $\sinh D \approx e^{D}/2$, and gives a long expression whose leading terms are
\begin{align}
\Sigma_2(\bar{\omega}) &\approx
- \frac{4}{9} Y(\bar{\omega}) e^{-2D_0} D_0^2
\big[D_0^2 + (16/3 - 4 \ln 2)D_0
\nonumber \\
&\quad + 52/3 + \pi^2/2 - 16\ln 2 + 6 \ln^2 2 \big].
\label{Sigma2}
\end{align}
The ratios $\Sigma_1/\Sigma_{\text{MB}}$ and $(\Sigma_1 + \Sigma_2)/\Sigma_{\text{MB}}$
are shown in fig.~\ref{fig:Ratios} as the green and red dashed curves. We see that $\Sigma_1 + \Sigma_2$ well approximates the exact result
in a wide range of $\bar{\omega}$.

\section{Discussion and conclusions}

All qualitative features of the functions $R$ and $S$ that we found, including the shape of the central 
peak and the features at the Mott scale, are in agreement with ref. \cite{Kirsch:2003}, see their eq. (4.20) and the list that follows. This is expected, since this paper deals with states deep in the tails of the spectrum, and so do we. However, when we compare our results with those obtained in the regime of large positive $E$ in refs. \cite{Gor'kov-I:1983, Gor'kov-II:1983} (conveniently summarized in tables I and II in \cite{Ivanov:2012}) we see many differences.

The values of the correlators $R$ and $S$ and their derivatives at $x=0$ differ between the two cases, but are not expected to be universal. The decay of the central peak $Q_0(x)$ has a different rate in the exponential, as well as a different power-law prefactor. The difference of the features at the Mott scale is more drastic: while in our case the behavior around $L_M$ is exponential (see eqs.~(\ref{Q_R-II}) and (\ref{Q_S-II})), it is an error function and a Gaussian for positive $E$. We attribute these differences to a different nature of the localized wave functions. The localized states in the optimal fluctuations of the disorder potential are simpler than the Anderson-localized states at $E \gg 0$. In particular, they do not have log-normally distributed tails, which, according to ref. \cite{Ivanov:2012}, affect the hybridization of wave functions at $E \gg 0$. In spite of this ``non-universality'' of the structure of the correlators $R$ and $S$ at the Mott scale, the behavior of the AC conductivity $\sigma(\omega)$ at low frequencies is still universal, and is given by the MB formula.

In conclusion, we have calculated exactly the Gaussian fluctuations around two-instanton saddle points of the functional integral
for the disorder average of two-point Green functions in a one-dimensional wire in the tail of the spectrum. This allowed us to derive the local DOS and dynamic correlation functions and the AC conductivity beyond the Mott-Berezinskii law, that is, for a broad range of frequencies. Unlike other approaches, our method can be applied to quasi-one dimensional systems~\cite{Gruzberg:2017}. We also hope that it can be useful for systems with non-Gaussian disorder-like random speckle potential~\cite{Falco:2010, Falco:2015}, and for systems with weak interactions~\cite{Falco:2009, Ivanov:2017}.

\acknowledgments

IG is grateful to D. Khmelnitskii for stimulating and illuminating discussions. AAF acknowledges support from the French Agence Nationale de la Recherche through Grants No. ANR-12-BS04-0007 (SemiTopo), No. ANR-13-JS04-0005 (ArtiQ), and No. ANR-14-ACHN-0031 (TopoDyn). IG was supported by the NSF Grant No. DMR-1508255.

\end{document}